# *In situ* Transmission Kikuchi Diffraction Tensile Testing

**Tijmen Vermeij[1*], Amit Sharma[1,2], Douglas Steinbach[3], Jun Lou[3], Johann Michler[1], Xavier Maeder[1]**

[1] *Laboratory for Mechanics of Materials and Nanostructures, EMPA, Thun, Switzerland*

[2] *Swiss Cluster AG, Spiez, Switzerland*

[3] *Department of Materials Science and NanoEngineering and the Rice Advanced Materials Institute, Rice University, Houston, TX, USA*

[*]Corresponding author: tijmen.vermeij@empa.ch

## Abstract

We present a methodology for *in situ* Transmission Kikuchi Diffraction (TKD) tensile testing that enables nanoscale characterization of the evolution of complex plasticity mechanisms. By integrating a modified *in situ* scanning electron microscope nanoindenter with a microscale push-to-pull device and a conventional Electron Backscatter Diffraction (EBSD) detector, we achieved TKD measurements at high spatial resolution during mechanical deformation. A dedicated focused ion beam procedure was developed for site-specific specimen fabrication, including lift-out, thinning, and shaping into a dog-bone geometry. The methodology was demonstrated on two case studies: (i) a metastable β-Ti single crystal, on which we quantified the initiation and evolution of nanoscale twinning and stress-induced martensitic transformation, and (ii) a CuAl/Al$_2$O$_3$ nanolaminate, which showed nanoscale plasticity and twinning/detwinning in a complex microstructure. Overall, this approach provides a robust alternative to *in situ* EBSD and transmission electron microscopy testing, facilitating detailed analysis of deformation mechanisms at the nanoscale. https://doi.org/10.1016/j.scriptamat.2025.116608

## Graphical Abstract

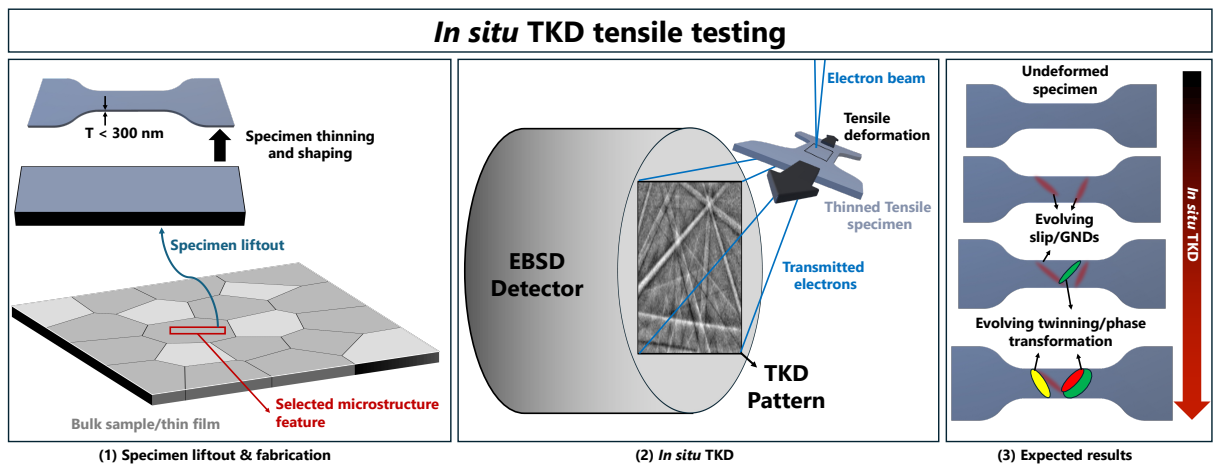

The advancement of the design of metal alloys, thin films and nanolaminates requires a deep understanding of nano- and microscale plastic deformation mechanisms. Traditional macroscale testing methods, even when complemented with post-mortem characterization or *in situ* characterization, fail to resolve fundamental deformation mechanisms in sufficient detail due to microstructural complexity and resolution limitations. Therefore, small-scale mechanical testing methods have emerged and are routinely employed inside Scanning Electron Microscopes (SEMs) to identify active crystallographic slip systems, quantify size effects, and measure fracture toughness at the microscale [1,2].

When microstructural features such as grain or phase boundaries are introduced in microscale specimens, full-field *in situ* characterization of deformation is required. Nanoscale Digital Image Correlation (DIC) can resolve crystallographic slip and strain partitioning at sufficient resolutions [3,4]. However, to quantify the evolution of deformation involving crystallographic events, diffraction-based measurements are required. Accordingly, Electron Backscatter Diffraction (EBSD) has been employed *in situ* on microscale specimens, e.g., (i) to resolve local stresses and Geometrically Necessary Dislocation (GND) density fields in micro-specimens [5–7], (ii) to measure the initiation and evolution of deformation twinning in hexagonal close-packed (HCP) metals [8,9], and (iii) to identify stress-induced martensite variants under deformation [10]. However, the spatial resolution of EBSD (~50 nm [11]) is insufficient to capture nanoscale deformation mechanisms in nanocrystalline materials and thin films.

In contrast, Transmission Electron Microscopy (TEM) can be used *in situ* with dedicated deformation stages to measure deformation at the nanoscale [12], even at atomic resolution [13]. While capable of imaging and diffraction-based orientation mapping over micrometer-sized areas [14], the integration of the diffraction signal over the specimen thickness can limit the spatial resolution for fine-grained specimens [15]. Additionally, TEMs are less accessible than SEMs and *in situ* TEM demands small deformation stages to fit within the TEM specimen port.

Alternatively, Transmission Kikuchi Diffraction (TKD) is an SEM-based technique in which an EBSD detector is employed to capture Kikuchi patterns in transmission, achieving spatial resolutions of 10 nm or better [16]. Notably, the field of view and in-plane specimen size are not limiting factors, and the specimen thickness requirements are less stringent than those for TEM [17]. As a result, TKD has been readily applied to characterize nanoscale deformation of various materials [18–22], and has even been used to measure the stress field around a single dislocation [23]. However, except for an initial exploration by Trimby *et al.* [24], TKD has not been applied *in situ* to measure deformations.

Therefore, this paper proposes a methodology for *in situ* TKD nano-tensile testing by integrating a modified *in situ* SEM nanoindenter with a microscale push-to-pull device and a conventional EBSD detector. We will describe the dedicated specimen fabrication routine and the subsequent application of *in situ* testing. The capabilities of the methodology will be demonstrated in two significantly different case studies. First, a metastable β-Ti single crystal will show nanoscale deformation twinning and stress-induced martensitic transformation. Second, *in situ* deformation of a nanolaminate thin film of sputter-coated crystalline CuAl and thin amorphous $Al_2O_3$ layers will confirm that the detailed deformation mechanisms of nanolaminates can be tracked.

The main technical challenge for realizing *in situ* TKD deformation involves the positioning of a sufficiently thin specimen correctly with respect to the EBSD detector while enabling the application of deformation. We considered fabricating a thinned tensile specimen near the edge of a bulk sample and applying deformation via a gripper, similar to the microscale *in situ* EBSD approach [8]. However, this method would require extensive Focused Ion Beam (FIB) milling - potentially using $Xe^+$ plasma FIB - and significantly limit flexibility in specimen selection. To overcome these challenges, we adopt the use of a micromechanical "push-to-pull" (PTP) device (see Figure 1(a)), which was originally designed to convert a compressive force from an *in situ* SEM/TEM nanoindenter into uniaxial tension for studying the mechanics of one-dimensional nanoscale samples, such as nanowires [25,26]. PTP devices have already been employed for *in situ* mechanical testing of metals within TEM and Transmission SEM (TSEM) environments [27,28]. Building on these prior works, we extend the application of the PTP device by integrating it with *in situ* TKD tensile testing.

The methodology is demonstrated using the metastable β-Ti single crystal case study. A PTP device [25] was mounted using silver paste on a dedicated holder that allows FIB preparation and SEM/TKD observation from multiple sides. Specimen lift-out, specimen transfer and specimen shaping mostly follow the procedures described by Stinville *et al.* [27]. Pt supports were created on the PTP device, see inset of Figure 1(a), to raise the specimen and to allow edge-on thinning with FIB. A ~25 μm length and ~1 μm thick FIB lift-out sample was extracted from a targeted grain (see Figure 1(b)), using a FIB-SEM dual beam microscope (Tescan Lyra 3). Next, the lift-out specimen was transferred to the PTP device (Figure 1(c)), where it was attached to both Pt supports using significant amounts of ion beam induced Pt deposition (Figure 1(d)). Before thinning, an initial dog-bone shape on the bottom side (non-Pt side) of the specimen was created using 30 kV & 500 pA FIB (Figure 1(e)). The Pt side of the specimen was retained until after thinning and polishing. Next, edge-on 30 kV FIB thinning was performed with stepwise milling steps, using ~1.5º over-tilt, with currents ranging from 500 pA down to 50 pA, resulting in a final specimen thickness of ~ 300 nm. The specific over-tilt angle was varied slightly between steps to minimize local variations in specimen thickness. Final polishing was performed using 5 kV & 50 pA FIB. The length of the thinned region was decreased stepwise with each thinning step, as shown in the final cross-section in Figure 1(f), to avoid severe changes in thickness that could lead to stress concentrations [27]. Finally, 30 kV 50 pA FIB milling was employed to create the dog-bone shape on the top side of the specimen. In this step, care was taken to avoid exposing the specimen surface to normal incidence ions, specifically by not taking a snapshot before cutting the gauge shape. Figure 1(g) shows the final tensile specimen, which has a gauge ~2 µm wide and ~3 µm long.

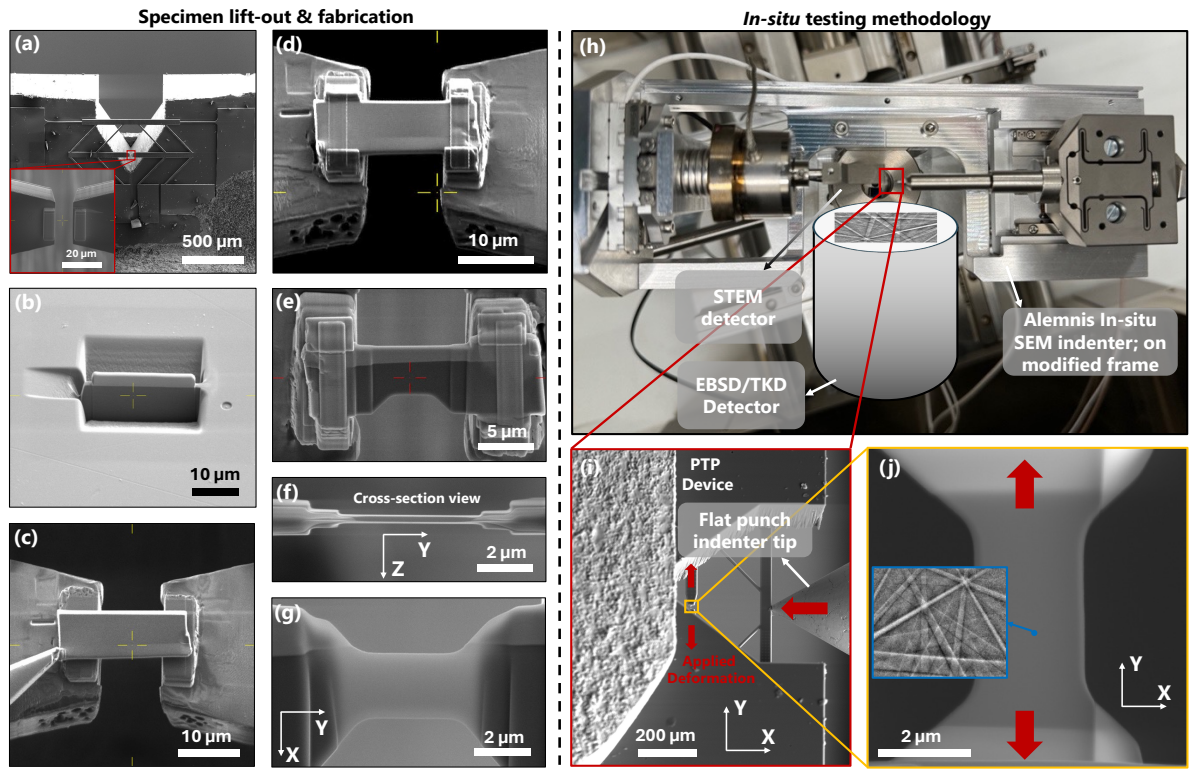

*Figure 1: Overview of the experimental methodology, with specimen lift-out and fabrication in (a-g) and in situ testing in (h-j). (a) PTP device with inset showing the FIB prepared specimen supports. (b-d) Specimen lift-out, transfer to PTP device and attachment by Pt deposition. (e) Specimen thinning and dog-bone gauge shaping. (f) Cross section view of thinned specimen. (g) Final specimen geometry after thinning and shaping. (h) Overview of Alemnis standard assembly nanoindenter, with in-house modified frame. EBSD detector, TSEM detector and relevant components of the indenter are indicated. The EBSD detector is only drawn schematically, with a representative experimental TKD pattern added, and does not represent the physical Symmetry I detector. (i) SEM view of PTP device in configuration for in situ TKD, with flat tip indenter coming in from the right. (j) SEM view of tensile specimen during in situ TKD, with a TKD pattern shown as inset. The TKD pattern is an average of 5 patterns, for illustrative purposes.*

Application of *in situ* TKD deformation was realized using an *in situ* nanoindenter (Alemnis AG, Switzerland), modified in-house with a custom frame (see Figure 1(h)) that allows (i) insertion of an EBSD detector in EBSD or off-axis TKD configuration without stage or frame tilt, (ii) operation of a modular TSEM detector beneath the PTP device, and (iii) usage of a small SEM Working Distance (WD) for optimal SEM/TKD measurements. The PTP device (Figure 1(i)), with prepared tensile specimen, is mounted upside down, under negative ~15º tilt, to assure that the transmitted electrons can reach the EBSD detector without any shadowing. A 15 µm flat punch indenter tip (Synton MDP) was employed to apply a compressive force to the PTP device, resulting in tensile deformation of the specimen, as illustrated in Figure 1(i). An Oxford Instruments Symmetry I EBSD detector was employed for off-axis TKD data collection, with the phosphor screen moved down 6 mm to achieve off-axis TKD patterns with optimal quality. All TKD data was indexed offline with spherical indexing using EMSphInx [29], for improved robustness to pattern quality degradation by plastic deformation. For EMSphInx, we used the pattern center parameters resulting from the calibration in the Oxford Instruments software. The *in situ* experiments were conducted inside a Tescan Lyra 3 dual-beam FIB-SEM, employing a 30 kV and 10 nA electron beam at 6 mm WD. *In situ* testing was only started after stabilization overnight to minimize drift. Additionally, TKD scan times were kept short, which also minimized contamination. Figure 1(j) shows a Secondary Electron (SE) image of the tensile specimen in position for *in situ* TKD, with a typical TKD pattern (averaged five times) shown in the inset. The indenter was operated in displacement control mode and the force signal was monitored for plasticity events, upon which the loading was paused and TKD mapping was conducted, after which deformation was continued. While force-displacement data can be retrieved, calculating stress-strain curves requires correction using finite element modelling [26], which is outside the scope of this work. Instead, we calculate the global engineering strain ($\varepsilon$) based on SEM images and include these values in the corresponding figures.

As a first case study, we investigate the deformation of a metastable β (BCC) Ti-12Mo sample, which is prone to deformation twinning and Stress-Induced Martensitic (SIM) transformation to orthorhombic $\alpha''$ [30,31]. A tensile specimen, selected to have a low maximum Schmid factor for crystallographic slip to promote twinning and martensitic transformation, was extracted, thinned to ~300 nm, and cut into a dog-bone shape as described above. *In situ* Bright-Field (BF) TSEM and TKD scans under tensile loading are shown in Figure 2, with the undeformed specimen depicted in Figure 2($a_1$, $b_1$). The TKD data, acquired on the full specimen with a step size of 20 nm with a scan time of ~20 min, is presented in Figure 2(b) by means of the Inverse Pole Figure (IPF) of the loading direction (y-direction). Both the BF TSEM and IPF maps show initial plasticity in the form of a single discrete slip band (see Figure 2($a_2$, $b_2$)), which progresses until new grains are detected near the slip band (Figure 2($a_3$, $b_3$)), leading up to specimen fracture, as seen in Figure 2($a_5$, $b_5$). Notably, the slip band does not propagate further upon initiation of twins and martensite variants, an observation only made possible by the *in situ* approach. Additionally, Figure 2($c_{2-5}$) shows first-neighbor Kernel Average Misorientation (KAM) maps (generated through MTEX), providing a qualitative indication of GND density. The initial slip band in ($c_2$) appears to evolve without a significant generation of GNDs. Upon twinning and phase transformation, complex dislocation structures appear alongside the deformation twinning and phase transformation.

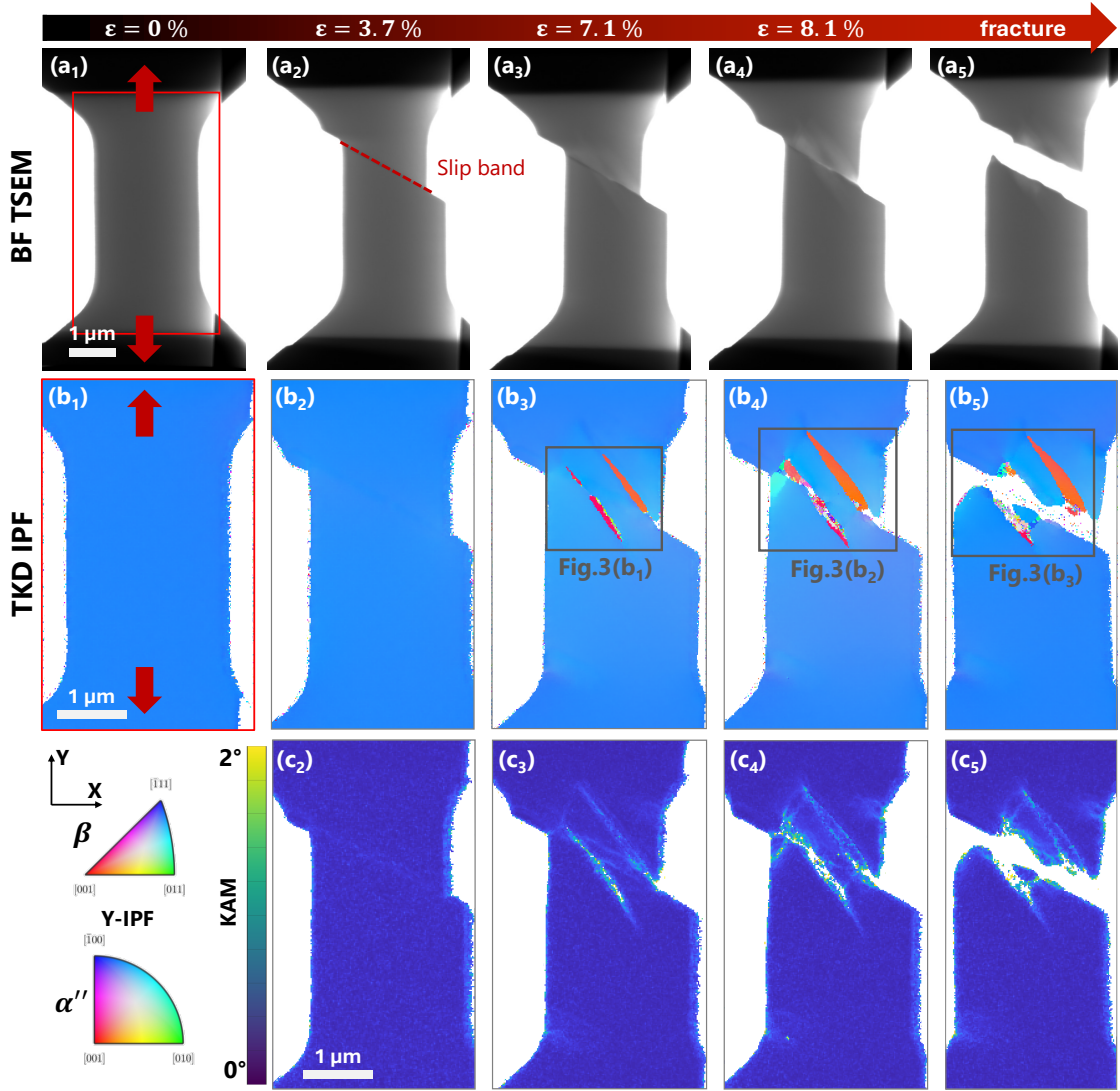

*Figure 2: In situ TKD deformation results of a β-Ti single crystal. (a$_{1-5}$) BF-TSEM images of the evolving deformation. (b$_{1-5}$) In situ TKD IPF maps (step size 20 nm), with areas indicated on which high-resolution scans were performed. (c$_{2-5}$) First-neighbor KAM maps, giving a qualitative indication of plasticity.*

Higher resolution TKD maps were acquired with a step size of 10 nm (scan time 40 min) in the most relevant areas and are shown in Figure 3. Using the MTEX toolbox [32], we identify the twinning systems and the crystallographic characteristics of the phase boundaries by assessing, for each grain boundary segment, if the measured misorientation conforms to a predefined twinning or martensitic Orientation Relationship (OR). Specifically, we consider (i) $\{332\}<113>_\beta$ and $\{112\}<111>_\beta$ deformation twinning within the β phase [31], (ii) the OR $\{\bar{1}10\}_\beta//(001)_{\alpha''} <111>_\beta//[110]_{\alpha''}$ [33] and $\{112\}_\beta//(110)_{\alpha''} <110>_\beta//[001]_{\alpha''}$ [34] for β/SIM$\alpha''$ phase boundaries, and (iii) $\{1\bar{3}0\}<310>_{\alpha''}$ deformation twins within SIM$\alpha''$ [31]. The results are shown as colored grain boundaries overlaid on the phase maps and IPF maps in Figure 3(a) and Figure 3(b), respectively. Initially, only $\{332\}<113>_\beta$ twins are observed, with a tiny double twin (width below 50 nm), as indicated in Figure 3(a$_1$, b$_1$). Upon further deformation, the twins grow, and small SIM$\alpha''$ martensite grains are formed (Figure 3($a_2$, $b_2$)), with several grains exhibiting ORs to both the parent grain and to the twins. Additionally, internal SIM$\alpha''$ $\{1\bar{3}0\} <310>_{\alpha''}$ twinning boundaries are observed. Figure 3($a_3$, $b_3$) shows the fractured state, revealing the growth and additional initiation of SIM$\alpha''$ variants. Notably, $\{112\}<111>_\beta$ twin boundaries were not observed for this specimen.

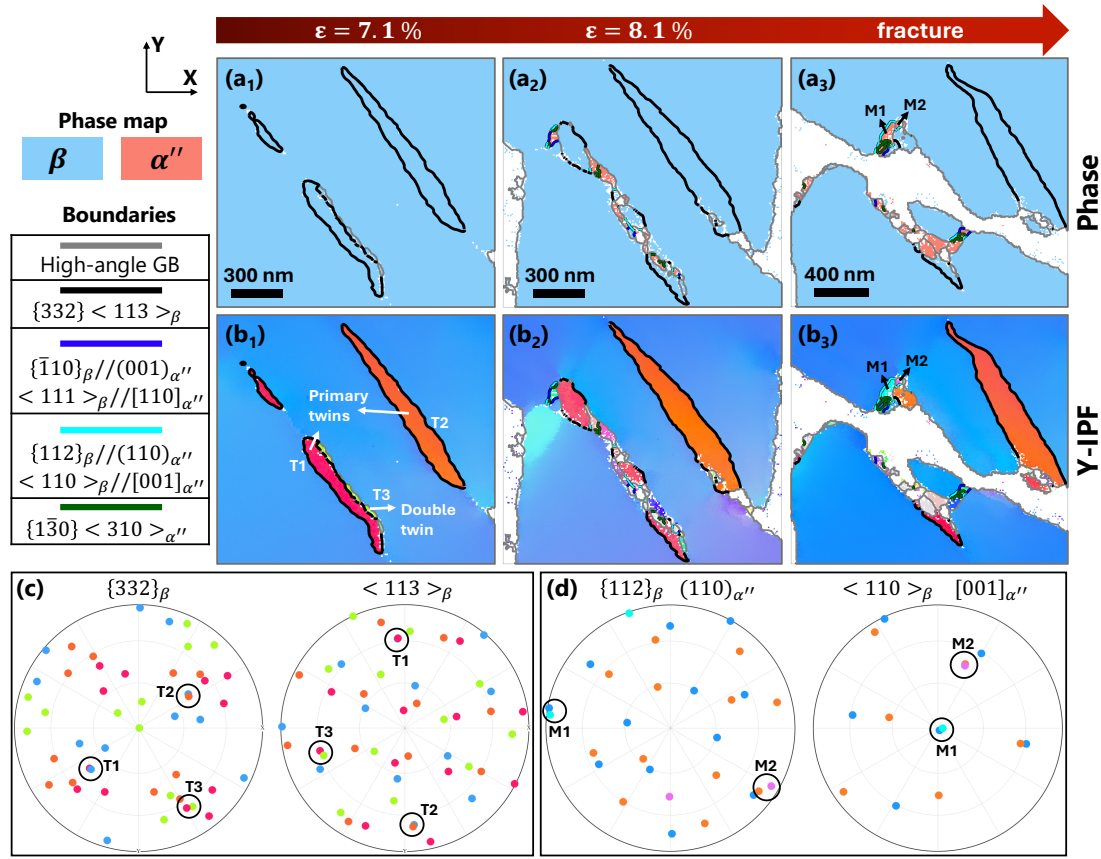

*Figure 3: High-resolution in situ TKD scans of the β-Ti single crystal (step size 10 nm) of the areas indicated in Figure 2($b_{3-5}$). ($a_{1-3}$) Phase maps and ($b_{1-3}$) Y-IPF maps with twin/phase boundaries colored according to the identified OR. For identification of $\{332\}<113>_\beta$ twinning a $4.5°$ threshold was used, according to Brandon's criteria [35]. For the ORs related to SIM$\alpha''$, thresholds of $10°$ were used. (c,d) Pole figure analyses of twinning and martensite ORs, color coded according to the IPF maps in (b). In (c), the $\{332\}<113>_\beta$ twins indicated in ($b_1$) are assessed. In (d), the $\{112\}_\beta//(110)_{\alpha''} <110>_\beta//[001]_{\alpha''}$ OR of SIM$\alpha''$ M1 and M2 variants as indicated in ($a_3$, $b_3$) is assessed.*

The misorientation-based identification of twinning and SIM$\alpha''$ ORs is confirmed with pole figures as shown in Figure 3(c,d). Figure 3(c) shows the aligned planes and directions of the twins indicated in Figure 3($b_1$), confirming the occurrence of the tiny double twin (T3). Figure 3(d) shows pole figures corresponding to the $\{112\}_\beta//(110)_{\alpha''} <110>_\beta//[001]_{\alpha''}$ β/SIMS $\alpha''$ OR, confirming the alignment of the planes and directions. This clearly shows that the $\alpha''$ M2 variant is related to a β twin.

The occurrence, sequence, and interaction of deformation twinning and SIM$\alpha''$ formation is discussed extensively in the literature, for a wide range of metastable β Ti alloys. Neutron and synchrotron-based

X-ray diffraction techniques are regularly employed *in situ* to monitor the formation of SIM$\alpha''$ in bulk β-Ti [36,37], but fail to capture the deformation mechanisms at microscopic scales. Alternatively, (quasi) *in situ* EBSD deformation experiments are performed on polycrystalline microstructures, usually followed by post-mortem TEM characterization [31,34,38,39]. In these studies, twinning and SIM$\alpha''$ transformation mostly initiate near pre-existing grain boundaries. Moreover, the grain size can determine the occurrence and sequence of the different mechanisms [38]. In contrast, our study eliminates the effect of grain boundaries and thereby provides a more fundamental platform for understanding SIM$\alpha''$ and twinning without pre-existing defects and under simple loading conditions. Due to the high spatial resolution of TKD, approaching that of TEM, we can show how the dynamic process of SIM$\alpha''$ formation occurs, revealing initiation of nanoscale SIM$\alpha''$ variants (<100 nm) near the twins, which could be easily missed by conventional EBSD. However, similar to (*in situ*) TEM studies, results should be interpreted with caution. Effects due to FIB damage (amorphization, Ga implantation, introduction of H) and electron beam damage must be considered. Additionally, the boundary conditions of a thin foil under deformation are not representative for bulk material, due to strong plane stress conditions [40] and free surfaces that can affect dislocation and twin nucleation [41].

Next, we transition from a single crystal to a complex nanolaminate. A 3 μm thick $Cu_{1-X}Al_X$ (X=0-12.5 at.%) multilayer film was deposited on a Si wafer using a novel SC-1 series deposition chamber (Swiss Cluster AG, Switzerland) that combines physical vapor deposition (PVD) and atomic layer deposition (ALD) without breaking the vacuum. Six 500 nm thick layers were deposited using PVD, starting with pure Cu, and increasing the Al content by 2.5 at.% for each subsequent layer, sequentially lowering the stacking fault energy [42]. Between layers, a 5 nm thick amorphous $Al_2O_3$ interface was deposited by ALD to prevent columnar grain growth and interlayer diffusion, and to improve mechanical properties [43]. The sample was then annealed at 500 °C for 1 hour in inert atmosphere (99.99% Argon), increasing the grain size to facilitate the application of TKD. A tensile specimen with a final thickness of ~150 nm was extracted and fabricated for *in situ* TKD, as shown in Figure 4(a).

BF-TSEM and TKD-based IPF and EMSphInx-based Cross-correlation Coefficient (CC) overlay maps of the CuAl-$Al_2O_3$ nanolaminate under increasing deformation are depicted in Figure 4($b_{1-5}$) and ($c_{1-5}$), respectively (step size 10 nm and scan time ~25 min per scan). Figure 4($d_{1-5}$) shows first-neighbor KAM maps, providing a qualitative indication of GND density, with overlay of grain boundaries and Σ3 twin boundaries corresponding to the $\{111\} < 112 >$ twinning system. Overall, the increasing levels of deformation are most apparent in the KAM maps. Only the final increment shows a significant difference in the TSEM and IPF maps (orange ovals in Figure 4($b_5,c_5$)). The final TSEM scan (Figure 4($b_5$)) shows overexposure of the BF TSEM due to extensive local thinning due to localized plasticity. The absence of a crack is confirmed by the presence of grains in the TKD-IPF maps (and by assessment of SE scans, not shown here). This shows that local thinning due to deformation does not significantly affect the TKD quality. The observed localized plasticity highlights the significant and remarkable ductility of the nanocrystalline microstructure, which is accommodated by the ductile thin amorphous $Al_2O_3$ layers [43,44].

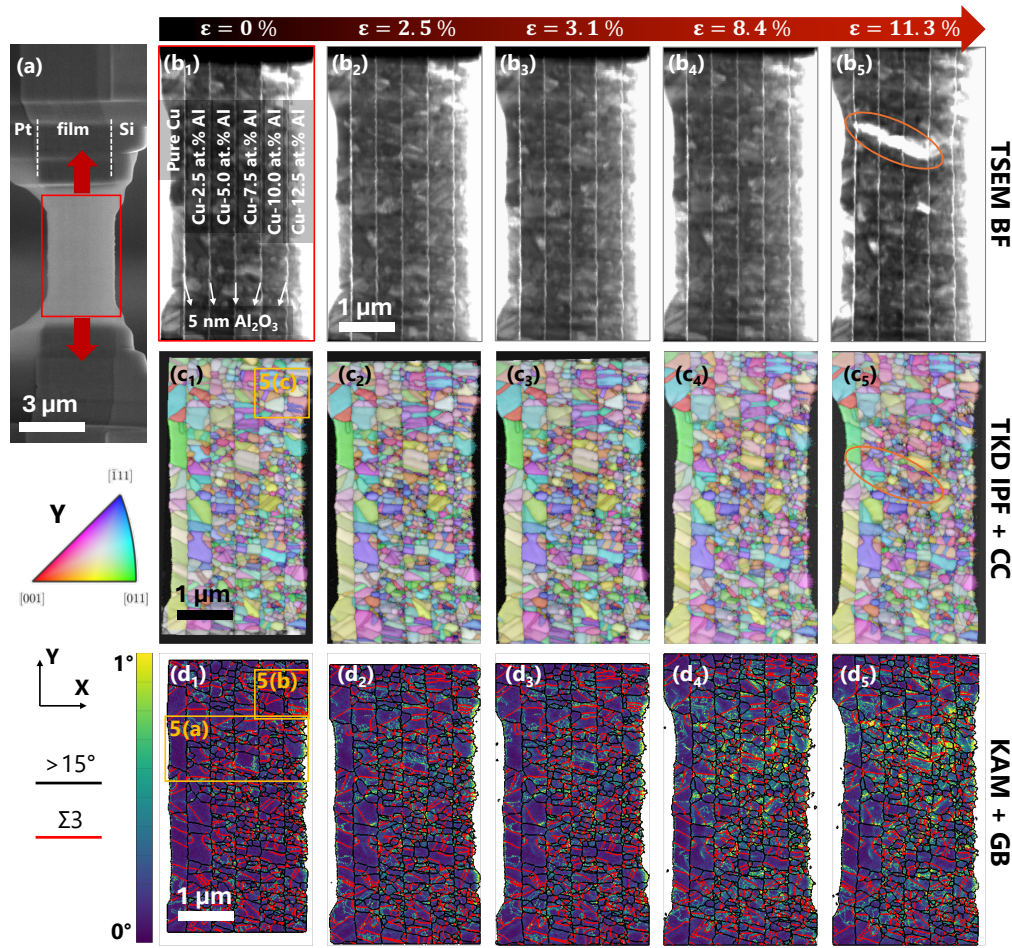

Figure 4: In situ TKD deformation of a CuAl-$Al_2O_3$ nanolaminate. (a) SE scan showing the tensile specimen geometry. ($b_{1-5}$) BF-TSEM scans over increasing deformation. ($c_{1-5}$) TKD-based IPF/CC maps over increasing deformation (step size 10 nm). ($d_{1-5}$) First-neighbor KAM maps with overlay of grain boundaries and $\Sigma 3$ twin boundaries. The orange oval indicates the localization area in both TSEM and IPF maps. Small areas, indicated by orange rectangles, are further analyzed in Figure 5.

Figure 5 shows two relevant areas at higher magnifications, as marked in Figure 4. Figure 5($a_{1-4}$) focuses on the area where plasticity localizes most, displaying the KAM maps which are qualitative indicators of GND density and thus primarily indicative of crystallographic slip. GND densities increase significantly with increasing global deformation, especially in the Al-containing layers. Even very small grains (<50 nm) exhibit significant dislocation activities, highlighting the capability of TKD to compete with TEM. While TEM integrates diffraction information over the entire specimen thickness, for TKD, the transmitting electron beam only produces diffraction information just before exiting the specimen. This implies that GNDs are captured primarily from the bottom of the lamella; however, it also reduces the likelihood of mis-indexing in cases where small grains do not span the full specimen thickness, as is likely the case here.

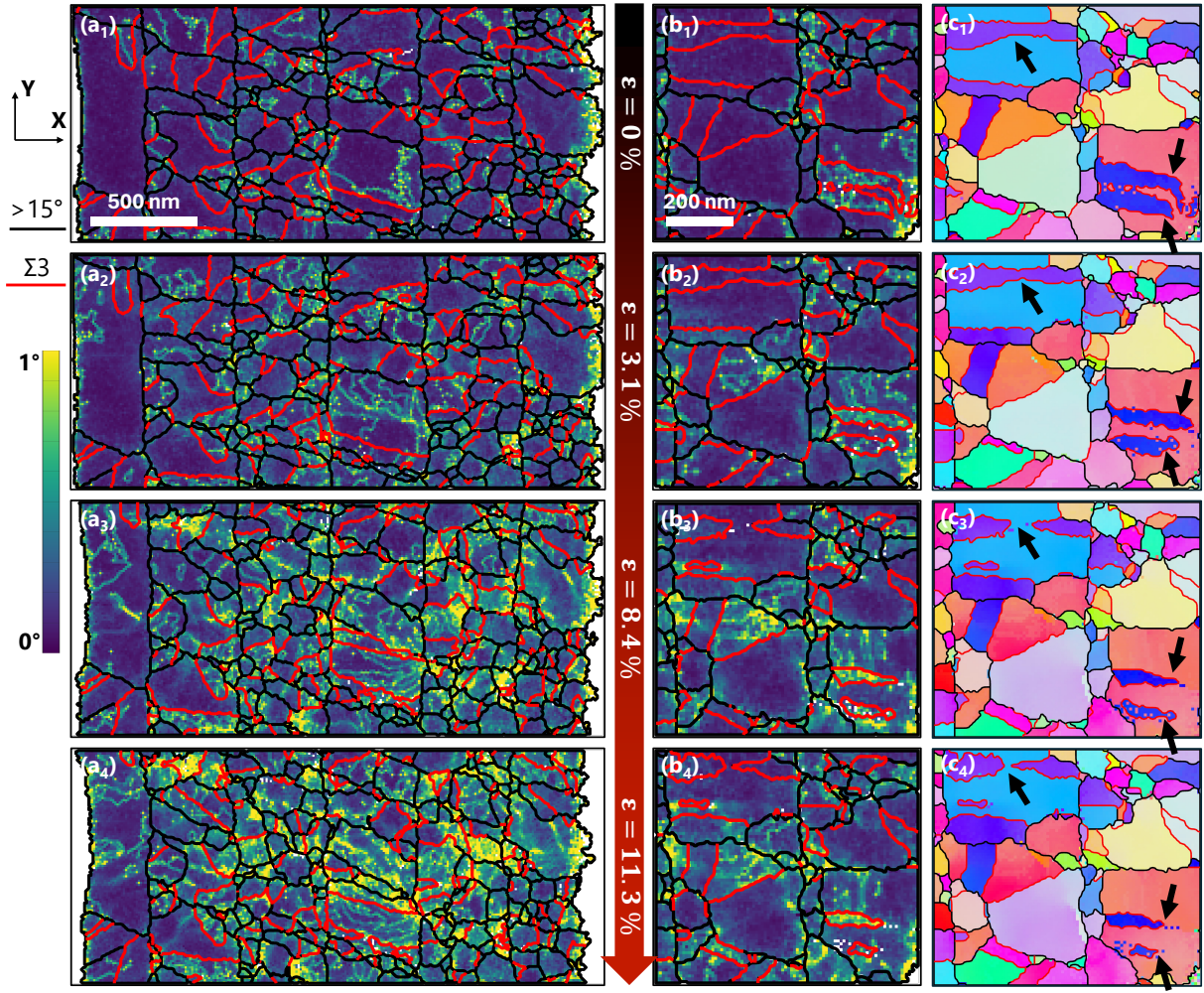

*Figure 5: Nanoscale deformation of CuAl-Al$_2$O$_3$ in two areas, magnified from the data of Figure 4. (a$_{1-4}$, b$_{1-4}$) First-neighbor KAM maps with GB overlay. (c$_{1-4}$) IPF maps with GB overlay. Detwinning is indicated by black arrows.*

While twin boundaries are present in each layer, there is no significant increase in the number of twin boundaries with increased levels of deformations, indicating that twinning does not contribute strongly to the plastic deformation. However, twin boundaries do move. Figure 5(b,c) shows that, with increasing plastic deformation, several twins reduce in size, as marked by black arrows in Figure 5(c), indicating that detwinning occurs. Indeed, detwinning has been shown to occur more readily than the initiation of new twins in similar microstructures [45,46]. Additionally, detwinning has been shown to be more prevalent for smaller grain sizes [47], a trend that is also reflected in our data, since the higher Al content films (films on the right) have finer grains.

In summary, we presented a methodology for *in situ* TKD nano-tensile testing that enables characterization of the evolution of various types of plasticity at the nanoscale. In a first case study on a metastable β-Ti single crystal specimen, the initiation and evolution of nanoscale twinning and stress-induced martensitic transformation were observed and quantified. A second case study, on a CuAl/Al$_2$O$_3$ nanolaminate, showed that nanoscale plasticity and twinning could be characterized in complex microstructures. These two case studies will be extensively investigated in future work. Furthermore, future application of *in situ* High angular Resolution TKD could yield measurements of local variations in stress [23]. However, the significant pattern distortion for off-axis TKD may require processing of the patterns with full-field correlation methods [48,49] and/or application of on-axis TKD [19]. Overall, the proposed methodology provides a solid compromise between *in situ* EBSD and *in situ* TEM testing.

## Author Contributions (CRediT)

**Tijmen Vermeij:** Conceptualization, Methodology, Investigation, Funding Acquisition, Writing-Original Draft, Visualization
**Amit Sharma:** Methodology, Investigation, Writing-Review & Editing
**Douglas Steinbach:** Investigation, Writing-Review & Editing
**Jun Lou:** Methodology, Writing-Review & Editing
**Johann Michler:** Resources, Writing-Review & Editing
**Xavier Maeder:** Conceptualization, Methodology, Investigation, Writing-Review & Editing

## Acknowledgements

The authors acknowledge Jérémie Bérard and Daniele Casari for technical and experimental support.

This work was financially supported by The Netherlands Organization for Scientific Research (NWO) by a NWO Rubicon grant, number 019.232EN.026. D.S. and J.L. would also like to acknowledge the support of The US Department of Energy, Office of Basic Energy Sciences under Grant Number DE-SC0018193.